\documentclass[bibyear]{aa}  
\usepackage{graphicx}
\usepackage{multirow}
\usepackage{txfonts}
\usepackage{natbib,twoopt}
\usepackage[breaklinks=true]{hyperref} 
\bibpunct{(}{)}{;}{a}{}{,}             
\newcommand{\customcite}[2]{\citeauthor{#1} (\citeyear{#1}; hereafter #2)}
\newcommand{\citecustom}[1]{K21}
\makeatletter
  \newcommandtwoopt{\citeads}[3][][]{\href{http://adsabs.harvard.edu/abs/#3}%
    {\def\hyper@linkstart##1##2{}%
     \let\hyper@linkend\@empty\citealp[#1][#2]{#3}}}
  \newcommandtwoopt{\citepads}[3][][]{\href{http://adsabs.harvard.edu/abs/#3}%
    {\def\hyper@linkstart##1##2{}%
     \let\hyper@linkend\@empty\citep[#1][#2]{#3}}}
  \newcommandtwoopt{\citetads}[3][][]{\href{http://adsabs.harvard.edu/abs/#3}%
    {\def\hyper@linkstart##1##2{}%
     \let\hyper@linkend\@empty\citet[#1][#2]{#3}}}
  \newcommandtwoopt{\citeyearads}[3][][]%
    {\href{http://adsabs.harvard.edu/abs/#3}
    {\def\hyper@linkstart##1##2{}%
     \let\hyper@linkend\@empty\citeyear[#1][#2]{#3}}}
\makeatother

\hypersetup{
    colorlinks=true,
    linkcolor=red, 
    citecolor=blue, 
    urlcolor=blue 
}
\def\mbh{$M_{\rm BH}$} 

\def\lbol{$L_{\rm bol}$}
\def\ledd{$\lambda_{\rm Edd}$}
\def\spin{$a^{\ast}$}
\def\fcol{$f_{\rm col}$}
\def\tlagobs{$\bar{\tau}_{\rm lag,obs}$}
\def\tlagmod{$\bar{\tau}_{\rm lag,mod}$}

\begin{document} 
   \title{X-ray reverberation modelling of the continuum, optical/UV time-lags in quasars}
   \author{D. A. Langis \inst{1,2}
\and
I. E. Papadakis\inst{1,2}
\and
E. Kammoun\inst{3,4}
\and
C. Panagiotou\inst{5}
\and
M. Dovčiak\inst{6}
}
\titlerunning{Time-lags of quasars.}
\authorrunning{D. A. Langis et al.}
\institute{Department of Physics, University of Crete, 71003 Heraklion, Greece \email{dlangis@physics.uoc.gr}
\and 
Institute of Astrophysics, FORTH, GR-71110 Heraklion, Greece 
\and
Dipartimento di Matematica e Fisica, Universit\`{a} Roma Tre, via della Vasca Navale 84, I-00146 Rome, Italy
\and
INAF -- Osservatorio Astrofisico di Arcetri, Largo Enrico Fermi 5, I-50125 Firenze, Italy
\and
MIT Kavli Institute for Astrophysics and Space Research, Massachusetts Institute of Technology, Cambridge, MA
02139, USA
\and
Astronomical Institute of the Academy of Sciences, Boční II 1401, CZ-14100 Prague, Czech Republic
}

\date{Received 5 May 2024 / Accepted 28 September 2024}

 
  \abstract
   { Extensive, multi-wavelength monitoring campaigns of nearby and higher redshift active galactic nuclei (AGN) have shown that the UV/optical variations are well correlated with time delays which increase with increasing wavelength. Such behaviour is expected in the context of the X-ray thermal reverberation of the accretion disc in AGN. }
   {Our main objective is to use time-lag measurements of luminous AGN and fit them with sophisticated X-ray reverberation time-lags models.  In this way we can investigate whether X-ray reverberation can indeed explain the observed continuum time lags, and whether time-lag measurements can be used to measure physical parameters such as the X-ray corona height and the spin of the black hole (BH) in these systems.} 
   {We use archival time-lag measurements for quasars from different surveys, and we compute their rest frame, mean time-lags spectrum. We fit the data with analytical X-ray reverberation models, using $\chi^2$ statistics, and fitting for both maximal and non spinning BHs, for various colour correction values and X-ray corona heights.}
   {We found that X-ray reverberation can explain very well the observed time lags, assuming the measured BH mass, accretion rate and X-ray luminosity of the quasars in the sample. 
   The model agrees well with the data both for non-rotating and maximally rotating BHs, as long as the corona height is larger than $\sim 40$ gravitational radii. This is in agreement with previous results which showed that X-ray reverberation can also explain the disc radius in micro-lensed quasars, for the same corona heights. The corona height we measure depends on the model assumption of a perfectly flat disc. More realistic disc models may result in lower heights for the X-ray corona. }
   
  

   \maketitle
%
\section{Introduction}
\label{sec:intro}

In the last decade, there have been many monitoring, multi-wavelength campaigns with the goal of studying the X-ray/UV/optical variations in active galactic nuclei (AGN). These campaigns are split, broadly speaking, into two categories. The first category includes the intensive monitoring of nearby AGN using the {\it Neil Gehrels Swift} Observatory, AstroSat (to a lesser extent), and in many cases ground based telescopes as well (e.g. \citealp{McHardy2014, Edelson2015, fausnaugh2016space, Edelson2017, cackett2018accretion, Chelouche2019, edelson2019, cackett2020supermassive, hernandez2020intensive, Vincentelli2021, kara2021agn, Donnan2023, miller2023continuum, kara2023uv, kumari2023contrasting, kumari2024detection}). Only a few, bright Seyferts belong to this category of intensively sampled observations over a prolonged period of time, albeit their number is increasing. The second category includes long-term monitoring campaigns of quasars, that is AGN which are at higher redshifts and are brighter than their nearby counterparts (e.g. \citealp{jiang2017detection, mudd2018quasar, homayouni2019sloan, yu2020quasar, guo2022active, jha2022accretion, homayouni2022sloan}). There are usually many AGN observed during each campaign, but their light curves are not as densely sampled as the ones from the monitoring of the nearby Seyfert galaxies. 

In all cases, the UV/optical variations are well correlated, with the variations at longer wavelengths being delayed with respect to the variations at shorter wavelengths. These results are consistent with the hypothesis of X-ray reverberation in AGN. According to this hypothesis, some of the photons emitted by the X-ray corona are directed straight to the observer, while the rest are directed towards the accretion disc. A fraction of the latter is ‘reflected' by the disc in the X-ray range. The remaining fraction is absorbed, increasing the temperature of the disc and the emitted UV/optical flux. Since X-rays are highly variable, the additional disc emission in the UV/optical bands is expected to be also variable, revealing a time lag (with respect to the X-ray), which should increase with wavelength (as observed). This phenomenon is referred to as the disc X-ray thermal reverberation (e.g. \citealp{kazanas01,cackett2007testing}).

In the last few years, we have studied in detail X-ray reverberation in AGN, for instance \customcite{kammoun2021uv}{K21a} and \cite{dovvciak2022physical}. We have been able to successfully fit the time-lags spectra of nearby AGN (\citealp{kammoun2019hard}, \customcite{kammoun2021modelling}{K21b}, \citealp{kammoun2023revisiting}), the mean and the variable, optical/UV/X-ray spectral energy distribution in NGC\,5548 (\citealp{dovvciak2022physical,kammoun2024broadband}), and the radial flux profiles in micro-lensed quasars (\citealp{papadakis2022x}). Additionally, \citet{panagiotou2020multiwavelength, panagiotou2022physical} utilised the same model to compute model power spectral density function (PSD) in the UV/optical range, and successfully fitted the optical/UV PSDs from the extensive monitoring campaign of NGC\,5548.

So far, we have studied results from the intensive monitoring campaigns of nearby Seyferts. In this work, we use archival time-lag measurements of high(er) redshift and more luminous AGN, compute their mean time-lags spectrum, and we fit them with the analytical models of K21a and b, considering the modifications recently proposed by \citet{kammoun2023revisiting}. Our objective is to investigate whether X-ray reverberation can explain the observed time lags in these objects as well.

Our paper is organised as follows: In Sect. \ref{sec:archival_tlags}, we present the archival time-lag measurements that we use in our sample. In Sect. \ref{sec:Models} we present the theoretical models we used, while in Sect. \ref{sec:fitthedata} we present the results of our analysis. Finally, we summarise and discuss the results and their implications in Sect. \ref{sec:Sum_Disc}.

\section{Archival time-lag measurements}\label{sec:archival_tlags}

\subsection{The data}
\label{sec:thedata}

We considered the time-lag measurements of \citet{jiang2017detection}, \citet{homayouni2019sloan}, \citet{guo2022active} and \citet{jha2022accretion}. Below we provide a short description of their data. 

\citet{homayouni2019sloan} studied quasar light curves from the Sloan Digital Sky Reverberation Mapping Project (SDSS-RM). They combined data from the SDSS-RM project and two ground based telescopes to construct light curves which were $\sim$200 days long in the \textit{g} and \textit{i} bands. They computed time lags between the two band light curves (with the \textit{g}-band as the reference) using both the interpolated cross-correlation function (ICCF; \citealp{gaskell1986line, gaskell1987accuracy, white1994comments, peterson2004black}) and JAVELIN\footnote{Whenever there are time lags computed using both ICCF and JAVELIN, we used the former. JAVELIN assumes a damped random walk model (DRW; \citealp{kelly2009variations, kelly2014flexible}) for the statistical description of the light curve variability. This model may not provide an accurate representation of the statistical properties of the variable processes in the optical/UV bands of AGN. For this reason, although both methods provide similar time-lag measurements, we chose ICCF based time lags.} (\citealp{zu2011alternative, zu2016application}). We used their ICCF time-lag measurements for their ‘well defined quasar sample' (95 objects), as listed in their Table 1. Redshift, monochromatic luminosity at 3000\AA~, $L_{3000\text{\AA}}$, and black hole (BH) mass, \mbh,~data were also taken from the same table. In order to compute the bolometric luminosity, \lbol, we multiplied $L_{3000\text{\AA}}$ by a factor of 5.15 (\citealp{richards2006spectral}). The time-lag measurements were not redshift corrected, so we multiplied them by $\frac{1}{1+z}$ (where \textit{z} is the redshift of each source), in order to bring them to the rest frame of each source. 

\citet{guo2022active} studied quasar light curves from the Zwicky Transient Facility survey (ZTF) in the \textit{g}, \textit{r} and \textit{i} bands, which were $\sim$1200 days long. They defined a core sample of 38 quasars, and we considered the ICCF time-lag measurements listed in their Table 3 for those targets. The time lags are given in the rest frame of each source.  Redshift, monochromatic luminosity at 5100\AA, $L_{5100\text{\AA}}$, and \mbh~are also listed in the same table. In order to compute \lbol, we multiplied $L_{5100\text{\AA}}$ by a factor of 9.26 (\citealp{richards2006spectral}).

\citet{jha2022accretion} studied quasar light curves obtained from the ZTF survey in the \textit{g}, \textit{r} and \textit{i} bands. They used both the JAVELIN and ICCF methods to compute the time lags between the \textit{g}-band and the \textit{r} and \textit{i}-band light curves. We considered the \textit{g}-\textit{r} and \textit{g}-\textit{i} ICCF time lags listed in their Table 3. There are 19 AGN with time-lag measurements in these bands, given in the rest frame of each source. Redshift, monochromatic luminosity at 5100\AA,~$L_{5100\text{\AA}}$, and BH mass for these objects were taken from their Table 1. As before, we multiplied $L_{5100\text{\AA}}$ by a factor of 9.26 in order to compute \lbol.

Finally, we also considered \citet{jiang2017detection}, who studied quasar light curves in the Medium Deep Fields (MDFs) of the Pan-STARRS1 survey (PS1), which were $\sim1200$ days long. They used JAVELIN to compute the time lags between the \textit{g}-band and the \textit{r}, \textit{i} and \textit{z}-band light curves. In this work we considered their \textit{g}-\textit{r}, \textit{g}-\textit{i} and \textit{g}-\textit{z} time lags, as listed in their Table 3, which were given in the rest frame of each source. There are 39 quasars with time-lag measurements in all bands. Redshift, bolometric luminosity, \lbol, and \mbh, for these quasars were taken from Table\,1 in \citet{jiang2017detection}. We note that, mixing of time-lag measurements from ICCF and JAVELIN could introduce a systematic error into the analysis, as these are fundamentally different methodologies. As we already mentioned, we chose to use time lags computed with the use of the ICCF method whenever possible, but \cite{jiang2017detection} do not use the ICCF method to compute time lags. \cite{homayouni2019sloan}, \cite{guo2022active}, and \cite{jha2022accretion} found that time lags computed by the two methods are broadly similar (see their Figs.\, 7, 3 and 4, respectively). This comparison implies that use of the \cite{jiang2017detection} time lags in our study will not bias our results significantly.


\begin{figure*}
    \centering
    \includegraphics[width=1.1\linewidth]{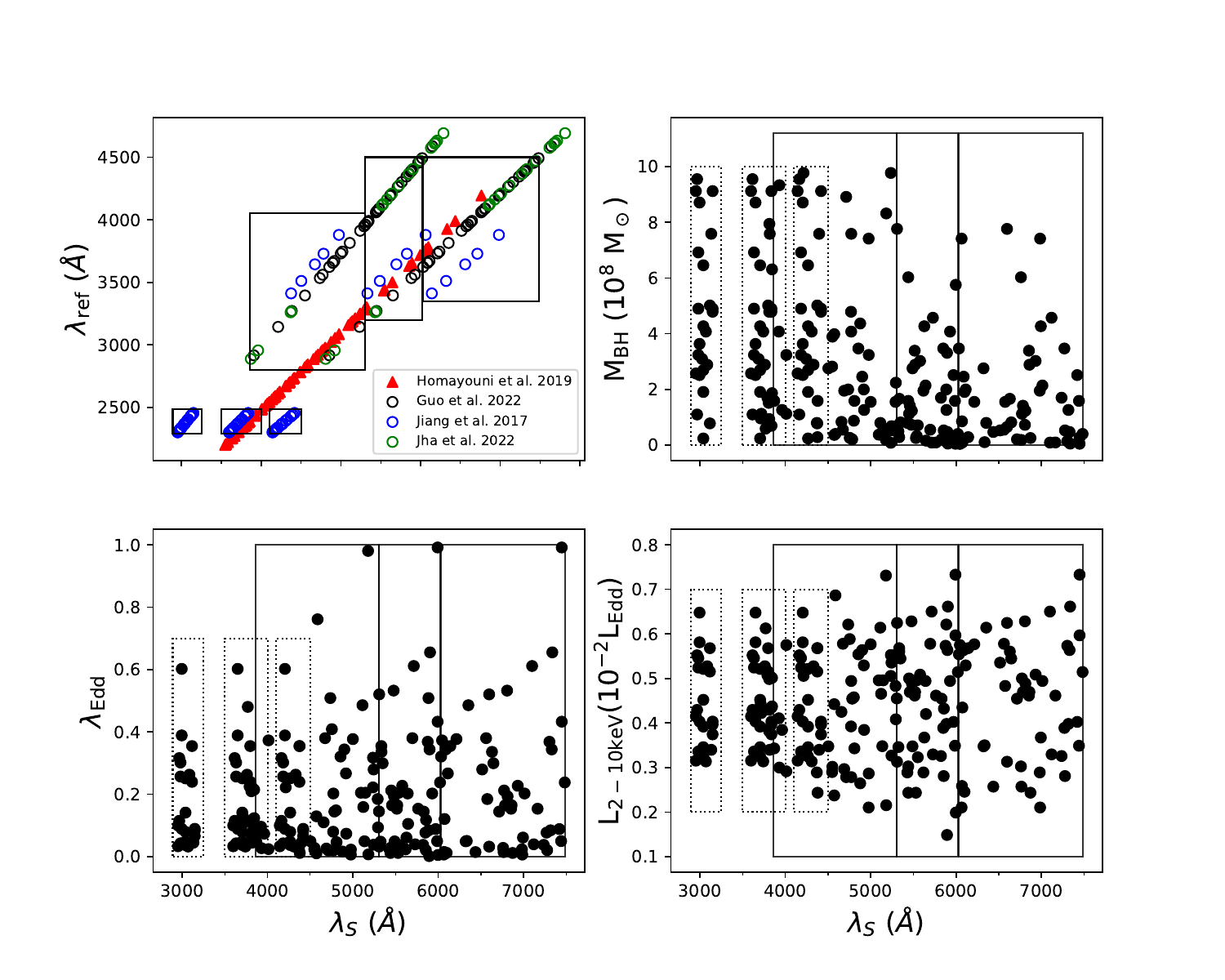}
    \caption{{\it Upper left panel}: The rest frame wavelength of the reference band (in our work always the $g-$band), $\lambda_{ref}$, versus the rest frame wavelength of the secondary light curves, $\lambda_{S}$, used to measure the time lags. {\it Upper right panel}: The \mbh\ as a function of $\lambda_S$ for the objects in our final sample. {\it Bottom panels}: Same as the top right panel but for \ledd\, and $L_{2-10 keV}$ (left and right panels, respectively). The boxes in all panels indicate the measurements we used to compute the mean time-lags spectrum.}
    \label{fig:refvssec}
\end{figure*}

In total, our initial sample consists of 326 time-lag measurements for 191 AGNe. All time lags are measured between the $r-$, $i-$ and $z-$band light curves and the $g-$band light curve, which is used as the reference band for all quasars in the sample.


\subsection{The observed time lags}

The top-left panel of Fig. \ref{fig:refvssec} shows the rest frame wavelength of the $g-$band light curves, $\lambda_{ref}$, plotted versus the rest frame wavelength of the secondary bands, $\lambda_{S}$. The sources have various redshifts, so $(\lambda_{\rm ref}$,$\lambda_S$) are not just a single point in this plot. Instead, three main diagonals appear in this figure, which correspond to the \textit{g-r}, \textit{g-i} and \textit{g-z} rest frame wavelengths.  There seems to be some scatter in the \textit{g-r} and \textit{g-i} diagonals, due to the fact that the effective wavelength of the respective filters in the PS1, SDSS-RM and the ZTF surveys does not appear to be the same in all of them. For the \citet{jiang2017detection} data we used the \textit{g}, \textit{r}, \textit{i} and \textit{z-}band effective wavelengths listed in \citet{tonry2012pan}. For the \citet{homayouni2019sloan} data we used the effective wavelengths in \citet{fukugita1996SDSSphot}. \citet{guo2022active} and \citet{jha2022accretion} used light curves from the ZTF survey and we assumed the effective wavelengths provided by the Spanish Virtual Observatory (SVO) Filter Profile Service.

The boxes in the top-left panel of Fig. \ref{fig:refvssec} indicate the data we use in our study. The data in the bottom left boxes correspond to sources with similar redshift. As a result, the spread of the rest frame $\lambda_{ref}$ and $\lambda_S$ is small for these objects. The top right boxes contain roughly the same number of time-lag measurements. Our objective is to calculate the average time lag in each box, and then compare the resulting mean time-lags spectrum with various time-lags models. 

The original sample of 191 AGN includes sources with a large range in \mbh\ and \lbol. As we explain in the next section, we use the analytical time-lags models of K21b, which were developed for AGN with $M_{BH}\le10^8M_\odot$ and accretion rate smaller than 0.5. For that reason, we kept in the sample sources with $M_{BH}<10^9M_\odot$ and \ledd\footnote{The Eddington ratio, \ledd, is defined as: \ledd=\lbol/$L_{\rm Edd}$, where $L_{\rm Edd}$ is the Eddington luminosity of the AGN. We compute \ledd\ using the \mbh\ and \lbol\ estimates of each source, and we assume it is a reasonable estimate of its accretion rate.}$\le 1$. If we accept lower limits on \mbh\ and \ledd\ then the number of time-lag measurements will decrease significantly  together with our ability to constrain models. In any case, we repeated the same analysis to smaller samples, with  (\mbh$<5\times 10^8$M$_{\odot}$, \ledd$<0.5$), as well as with (\mbh$<2\times 10^8$M$_{\odot}$, \ledd$<0.5$). In both cases, the results are almost identical, although with larger errors.

\begin{figure}[!htbp]
    \centering
    \includegraphics[width=\linewidth]{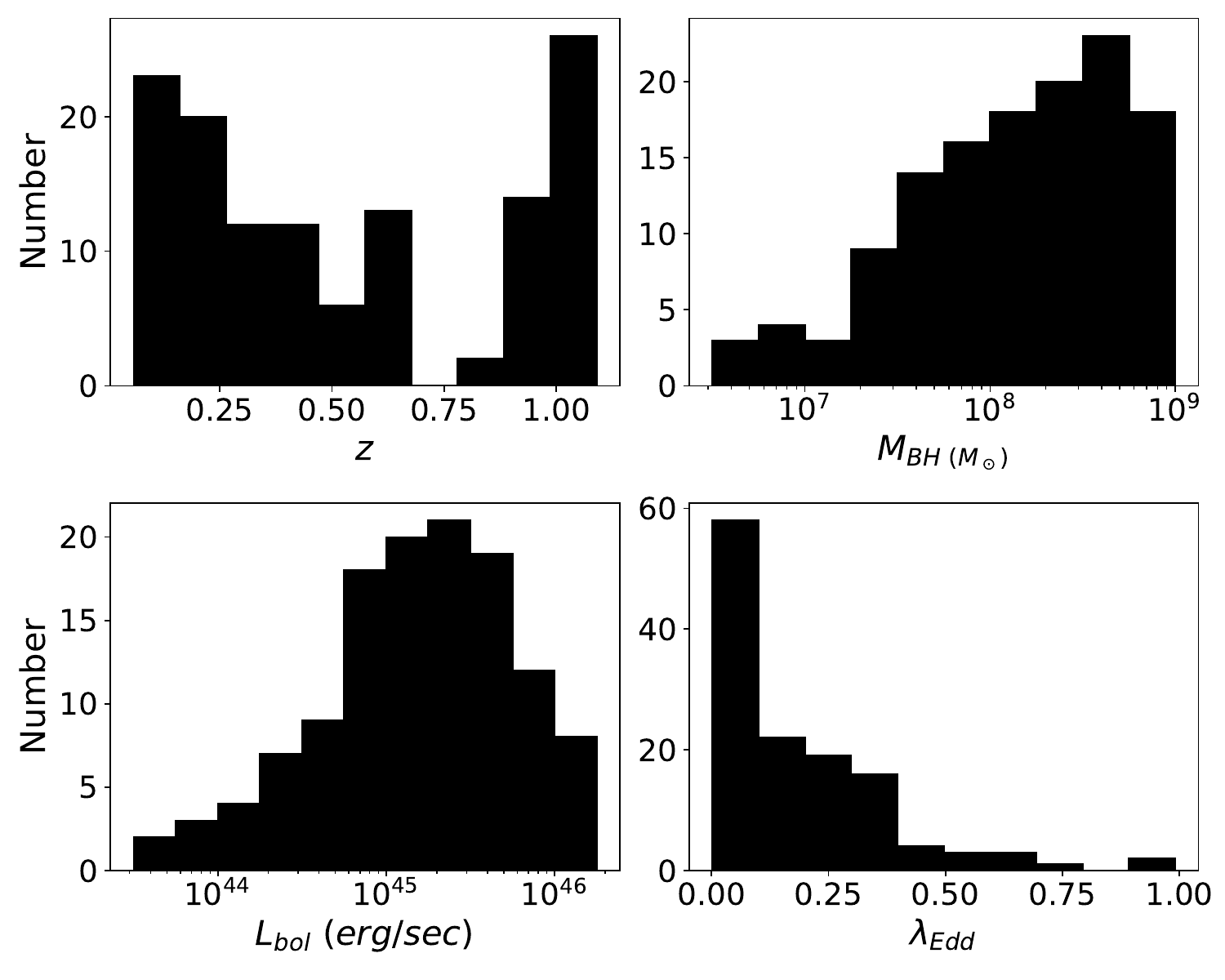}
    \caption{The redshift, \mbh, \lbol\ and \ledd\ distributions for the sources in the final sample.}
    \label{fig:distr}
\end{figure}

Due to the reasons we discussed above, we finally kept 128 objects in the final sample.  Figure \ref{fig:distr} shows the distribution of their \textit{z}, \mbh, \lbol~and \ledd. Redshifts range from $\sim0.05$ to $\sim1.1$. Most of the quasars in the sample have redshifts smaller than $\sim 0.45$, but there is also a large number of objects with redshift larger than $\sim 0.8$. This bi-modal distribution is due to selection effects. We note that our sample is not a complete sample of quasars. It simply consists of sources with measured continuum time lags in the UV/optical bands. Nevertheless, their properties are representative of quasars in larger samples. 
For example, the upper right plot in Fig.\,\ref{fig:distr} shows the distribution for the BH mass, and the bottom left plot shows the distribution for \lbol. The BH mass is larger than $10^8M_\odot$, and \lbol\ is larger than $10^{45}$ ergs/s for most sources. The range of the \mbh\ and \lbol\ values for the sources in our sample is within the range of the respective distributions in other quasar samples like, for example the SDSS quasars in Stripe 82 \citep[see e.g.][]{Petrecca24}. We are missing quasars with BH mass larger than $\sim 10^9$ M$_{\odot}$ and \lbol$>10^{46}$ erg/s, as the optical/UV continuum time lags have not been measured yet in such objects. On the other hand,  the sources we study are brighter, and host BHs which are more massive than the BHs in nearby AGN \cite[see e.g.][]{koss2022bass}. The bottom right plot in the same figure shows the distribution for the \ledd, which ranges from $\sim 1.5\times 10^{-3}$ up to $1$. The accretion rate is smaller than $\sim 0.4$ for the majority of the sources in the sample, which is comparable to the accretion rate of AGN in the nearby Universe \citep{koss2022bass}.

\subsection{The observed, mean time-lags spectrum}
\label{sec:meanobstimelags}

\begin{figure}[t]
    \centering
    \includegraphics[width=\linewidth]{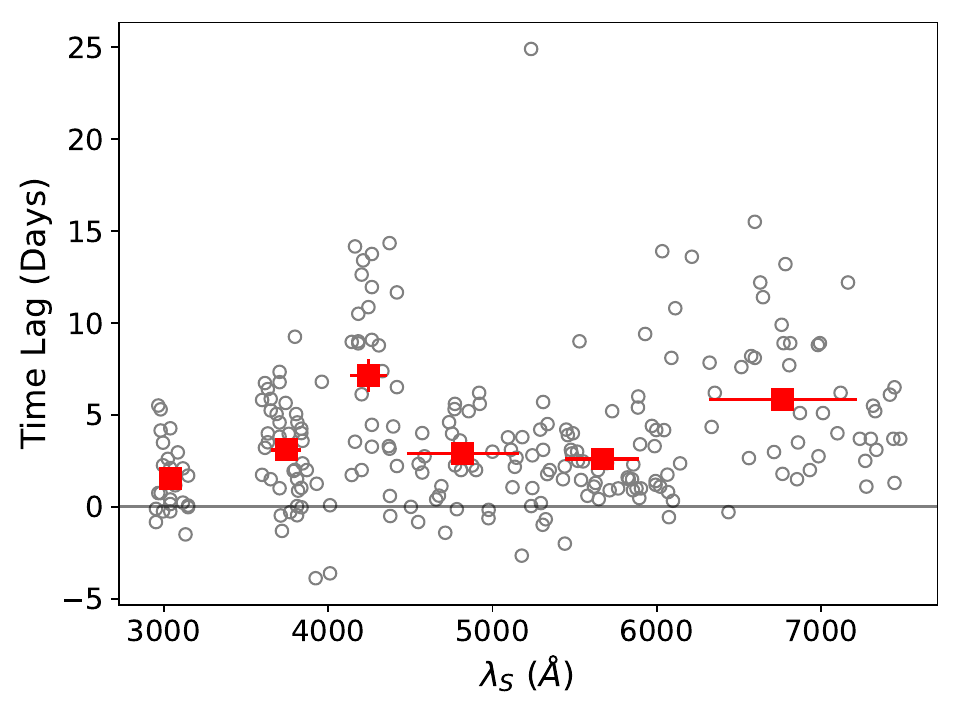}
    \caption{The observed time lags as a function of $\lambda_{S}$. Red filled circles show the mean time lags (computed as explained in Sect. \ref{sec:meanobstimelags}). }
    \label{fig:obstlags}
\end{figure}

The grey open circles in Fig. \ref{fig:obstlags} indicate all time lags for the AGN in the final sample (231 measurements in total). The observed time-lags spectrum is very scattered, mainly because each point in this figure corresponds to the time lag between various wavelengths, for quasars with different \mbh, \ledd, and X--ray luminosity.  This is shown in Fig.\,\ref{fig:refvssec}. The upper right and bottom panels in this figure show the BH mass, accretion rate and X--ray luminosity of the sources in the sample, plotted as functions of $\lambda_S$. Data appear in roughly six groups in these plots, which correspond to the six boxes we defined in the top-left panel in the same figure. 

The max-to-min ratio of \mbh\, and \ledd\ is larger than $\sim 20$ in some cases. Both of these parameters affect the magnitude of the time lag, hence the large spread shown in the relevant panels in Fig.\,\ref{fig:refvssec} can account, to a large extent, for the spread of the observed time lags. The bottom right panel shows the spread of the 2-10 keV luminosity as well (computed as explained in Sect.\, \ref{sec:fitthedata}). The X-ray luminosity also affects the magnitude of the time lag in the optical/UV bands (see e.g. Fig. 23 in K21a). Therefore, part of the scatter of the time lags plotted in Fig. \ref{fig:obstlags} is also due to the spread of the $L_{\rm 2-10 keV}$ values that is shown in this panel. In addition, in some cases, the rest frame wavelength of the reference band, $\lambda_{ref}$, is not the same for a given $\lambda_S$: time lags at the same $\lambda_S$ indicate delays between the observed variations in this wavelength and the variations in different $\lambda_{ref}$'s, depending on the redshift of each source (this is the case for sources which belong in the second and third lower boxes and in the first upper box in the top-left panel of Fig.\,\ref{fig:refvssec}). 

\begin{figure}
    \includegraphics[width=1\linewidth]{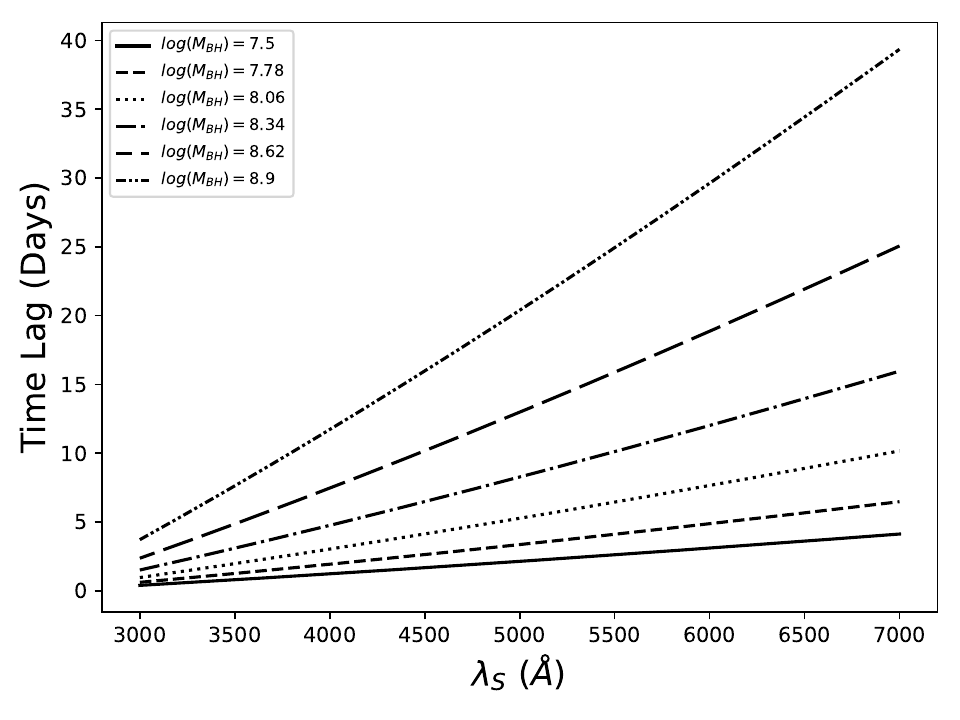}
    \includegraphics[width=1\linewidth]{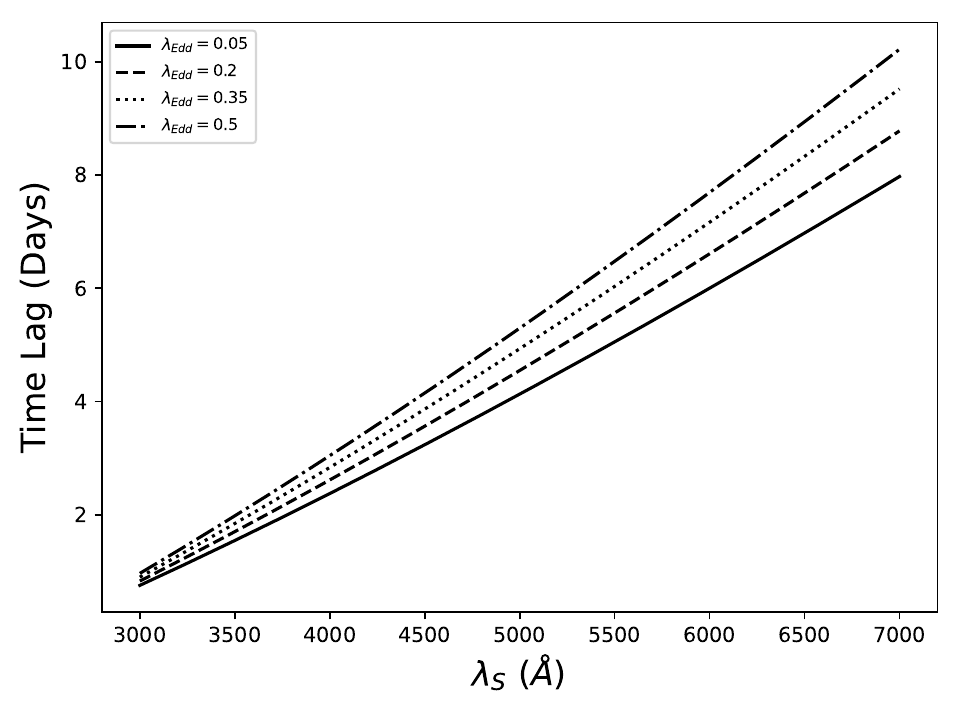}
    \includegraphics[width=1\linewidth]{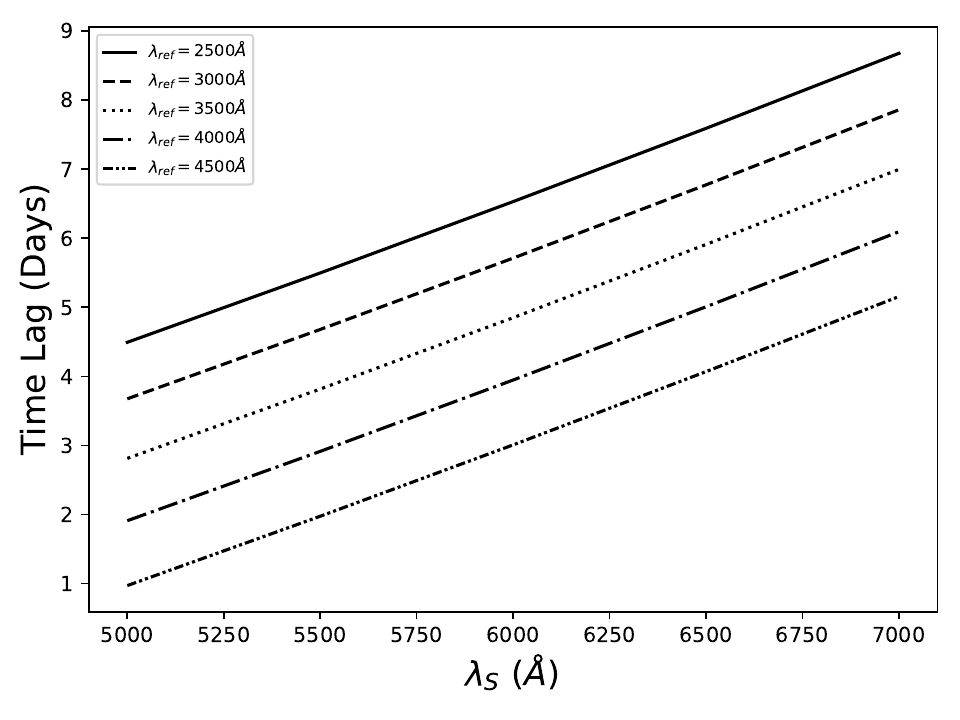}
    \caption{The dependence of the time lags on BH mass, accretion rate and reference wavelength (top, middle and lower panels, respectively), according to the K21a,b time-lags models (see text for details). }
    \label{fig:modeldependence}
\end{figure}

Given all the above, modelling of the observed time lags plotted in Fig.\,\ref{fig:obstlags} is not straightforward. 
In order to fit the data, we must use models which take into account the dependence of time lags on the BH mass, accretion rate, X--ray luminosity and the difference between the reference and the second wavelength. This is the case of the X-ray reverberation models of K21a and K21b. To demonstrate this, Fig.\,\ref{fig:modeldependence} shows the model X-ray reverberation time lags for various BH masses, accretion rates and reference wavelengths (top, middle and bottom panels, respectively). The time lags have been computed using the equations in Section 4 of K21b, for a non-rotating BH and a corona height of 20 R$_{g}$. The reference wavelengths in the top and middle panels is $\lambda_{ref}=2500$\AA. 

The top panel in Fig.\,\ref{fig:modeldependence} shows a plot of the model time lags for BH masses between $\sim 3.2\times 10^7$M$_{\odot}$ and $\sim 8\times 10^8$M$_{\odot}$ (this is roughly the BH mass range for  most of the sources in our sample; see top right panel in Fig.\,\ref{fig:distr}). We have assumed an accretion rate of 0.18 in all cases (which is the mean accretion rate of the quasars in our sample). Clearly, we  expect a large scatter in the observed time lags, if we consider a sample of sources with a large range of BH masses. Even if the BH mass is fixed, we still expect a (smaller) scatter in the observed time lags, if we study objects with a range of accretion rates (for reasons explained in  Section 3.2 in K21a; see also their Fig. 17). This is shown by the middle panel in Fig.\,\ref{fig:modeldependence} (time lags in this figure are computed for an AGN with a BH mass of 2.6$\times 10^8$M$_{\odot}$, which is similar to the mean BH mass of the objects in the sample). The range of accretion rates we have considered is representative of the range of the \ledd\ estimates in the final sample (see the bottom right panel in Fig.\,\ref{fig:distr}). Clearly, a range of accretion rates will also contribute to the scatter of the observed time lags (albeit by a smaller factor, when compared to the scatter induced by the difference in BH mass). 

The bottom panel in Fig.\,\ref{fig:modeldependence} shows the expected scatter in the observed time lags due to the fact that, in some cases, the time lags at a given wavelength, $\lambda_{S}$, indicate delays between $\lambda_{S}$ and various $\lambda_{ref}$. The model time lags in this panel are computed for \mbh=2.6$\times 10^8$M$_{\odot}$ and \ledd=0.18, while the range of reference wavelengths that we have considered is representative of the range of the reference wavelengths in the sample (see top left panel in Fig.\,\ref{fig:refvssec}). 

The discussion above shows that X-ray reverberation can explain, qualitatively, the large scatter of the observed time lags in Fig.\,\ref{fig:obstlags}. However, it is necessary to also fit the model to the data to investigate whether X-ray reverberation can actually fit the observed time lags. It is not straightforward how this can be done, given the fact that the data plotted in Fig.\,\ref{fig:obstlags} depend on various model parameters. One possibility would be to fit X-ray reverberation time-lags models to each individual source plotted in Fig.\,\ref{fig:obstlags}. However, individual time lags have large errors, which in many cases appear to be highly asymmetric (as reported in the literature). 

Another possibility would be to compute the mean time lag, \tlagobs, and the mean secondary wavelength, $\bar{\lambda}_S$, in each of the six boxes plotted in the top-left panel in Fig.\,\ref{fig:refvssec}. Since we can predict the theoretical model time lag for each source (assuming their BH mass, accretion rate, and X-ray luminosity), we could also compute the mean model time lags in the same way we compute the mean observed time lags, and then fit the mean model to the mean observed time lags. The advantage of this approach is that, by averaging many measurements, the error of the mean time lag will be significantly smaller than the error of each individual time-lag measurement and, at the same time, the mean will be approximately a Gaussian variable. As a result, we will be able to use traditional $\chi^2$ statistics to fit the model to the data. 


The red-filled squares in Fig. \ref{fig:obstlags} show \tlagobs\ plotted as a function of $\bar{\lambda}_S$. 
The $y-$axis error indicates the standard error of the mean, while the horizontal ‘error-bar' indicates the scatter of the secondary wavelengths around $\bar{\lambda}_S$. The \tlagobs($\bar{\lambda}_S$) points in this figure define the observed, mean time-lags spectrum of the quasars in the sample, and we explain below how we fit it with the model X-ray reverberation time lags. We stress that the term ‘mean time-lags spectrum' does not imply that the red points in Fig. \ref{fig:obstlags} indicate the intrinsic, true mean time-lags spectrum of AGN as a whole. Instead, this is the ‘mean' time-lags spectrum of this particular sample of objects (which, as we commented above, is not a complete sample of AGN), which was computed using time-lag measurements at various filters, as reported by the papers we considered.

\section{Theoretical models}
\label{sec:Models}

The X-ray reverberation time-lags models of K21a and K21b were based on the disc response functions of K21a. These functions were computed assuming an X-ray point like source illuminating a standard Novikov-Thorne (NT) accretion disc \cite[]{novikov1973astrophysics}. The X-ray source is located at a specified height, \textit{h}, along the rotational axis of the BH and emits, isotropically in its rest frame, a power law spectrum with a photon index of $\Gamma$. Part of the X-ray flux that is received by the disc is reflected and re-emitted in X-rays, and the rest is absorbed. The absorbed X-rays act as an additional source of heating for the disc, leading to an increase of the local disc temperature and emission. Since the X-rays are variable, the extra disc emission should also be variable. X-ray reverberation predicts delays between the optical/UV variations, which should increase with increasing wavelength, since the X-rays initially illuminate the inner regions of the disc before propagating outwards, to the outer disc region. K21a computed model time lags, $\tau_{\rm mod}(\lambda)$, according to the relation,

\begin{equation}
    \tau_{\rm mod}(\lambda)=\frac{\int t\Psi(t,\lambda)dt}{\int\Psi(t,\lambda)dt},
    \label{eq:tlagmod}
\end{equation}

\noindent where $\Psi(t,\lambda)$ is the disc response function, which is defined as follows: 

\begin{equation}\label{eq:responsefunc}
    \Psi(t,\lambda)\propto\frac{F_{tot}(\lambda,t_{obs})-F_{NT},(\lambda)}{L_X}.
\end{equation}

\noindent where $F_{\rm tot}(\lambda, t_{obs})$ in the equation above is the disc flux at wavelength $\lambda$ that a distant observer will observe at time $t_{\rm obs}$, while $F_{NT}(\lambda)$ is the NT disc flux at the same wavelength. In effect, the disc response function is the extra, time variable disc flux due to the X-rays that are absorbed by the disc. The response functions were calculated taking into account all the relativistic effects in the propagation of light from the corona to the disc, and from the disc to the distant observer. K21a also considered the disc ionisation state in order to calculate as accurately as possible the X-ray flux that is reflected by the disc. 

K21a computed model time lags using eq.\,\eqref{eq:tlagmod}, and derived an analytic prescription for the time lags between optical/UV and X-rays. K21b slightly modified the K21a results, and provided an analytic model for the time lags between two wavelengths, $\lambda_{\rm ref}$ and $\lambda_{S}$, within the optical/UV bands. We use their models to fit the mean time-lags spectrum of the quasars in our sample. The main physical parameters of the model are the BH mass (in $10^7M_\odot$), the accretion rate (in 0.05 of the Eddington limit), the height of the corona (in $10R_g$), and the 2–10keV luminosity (in units of 0.01 of the Eddington luminosity). Given these parameters, the model can predict the time lag between the disc emission at wavelengths $\lambda_{\rm ref}$ and $\lambda_S$ within the UV/optical band, for a maximally or non spinning BH (i.e. \spin=0\footnote{The dimensionless spin parameter is defined as \spin=$Jc/G$\mbh$^2$, where $J$ is the angular momentum of the BH. It is smaller than or equal to 1.} and \spin=1, respectively). 

The K21a time lags were computed assuming a colour correction factor of \fcol=2.4. Colour correction factors are frequently used in order to take into account the effect of electron scattering of the disc photons, which may play a significant role and could lead to deviations from blackbody emission. 
A colour correction factor of $f_{col}=2.4$ is appropriate in the case of AGN with a large accretion rate \cite[]{ross1992spectra}. Recently, \citet{kammoun2023revisiting}, provided an analytic correction for the K21a and K21b models, for any \fcol\ value (see their Appendix B1).

\begin{center}
\begin{table}[]
\centering
\caption{Best-fit results.}
\label{tab:results}
\begin{tabular}{llccc}
\hline
Spin                          & $f_{col}$(Model) & \begin{tabular}[c]{@{}c@{}}Best fit height\\ ($R_g$)\end{tabular} &  $\chi^2_{\rm min}$/dof \\ \hline
\multirow{3}{*}{\spin$=0$} & 1(M01)         & 80    & 80.67/5         \\
                              & 1.7(M01.7)       & 42$^{+34}_{-23}$  & 11.4/5          \\
                              & 2.4(M02.4)       & 5         & 30.1/5         \\ \hline
\multirow{3}{*}{\spin$=1$} & 1(M11)         & 80     & 124.7/5       \\
                              & 1.7(M11.7)       & 80    & 33.6/5       \\
                              & 2.4(M12.4)       & 50$^{+15}_{-12}$   & 11.3/5     \\\hline
\end{tabular}
\end{table}
\end{center}

\section{Fitting the observed time lags}
\label{sec:fitthedata}

We fitted the observed mean time-lags spectrum assuming six different X-ray reverberation time-lags models with \spin$=0$ and $f_{col}=1,1.7, 2.4$ (models M01, M01.7, and M02.4, respectively), as well as \spin$=1$ and $f_{col}=1,1.7,2.4$ (models M11, M11.7, and M12.4). To compute the time lags when \spin$=0$ we use eqs. (2)-(5) in K21b, while for \spin$=1$ we use their eqs. (6)-(9). In order to compute the time lags for the $f_{col}=1$ and $f_{col}=1.7$, we added the term for the colour correction prescription, which is listed in eq. (B1) and eq. (B2) in Appendix B1 in \citet{kammoun2023revisiting}.

First we computed the model time lags for each of the sources in the boxes plotted in the top-left panel of Fig.\,\ref{fig:refvssec}. We used the \mbh\, and \lbol\ values from the literature (taken as described in Sect. \ref{sec:thedata}), and we computed \ledd. We also computed the 2–10 keV luminosity assuming that log(\lbol/$L_{\rm (2-10) keV}) = mlog($\ledd$)+q$, where the $m=0.75\pm 0.04$ and $q=2.13\pm0.04$ \citep{lusso2012bolometric}. Knowing \mbh, \ledd, \spin, and \fcol, the only free parameter of the time-lags models is the X-ray corona height, $h$. For each source, we computed model time lags for corona heights in the range between 5 and 80 $R_g$, with a step of 1 $R_g$\footnote{$R_g=G M_{\rm BH}/c^2$, is the gravitational radius.}. 

Once we computed time lags for the individual sources then, for each height, we computed the mean model time-lags spectrum in each box, \tlagmod($\bar{\lambda}_S,h)$, and the $\chi^2$ statistic, that is $\chi^2(h)=\sum_{i} [$\tlagobs$(\bar{\lambda}_{S,i}) - $\tlagmod$(\bar{\lambda}_{S,i},h)]^2/\sigma^2($\tlagobs$_{,i}$), where $i=1,\dots,6$ (i.e. the number of boxes where we computed the observed mean time lags). The height which gives the minimum $\chi^2$ ($\chi^2_{\rm min}$) is the best-fit model for each of the M01, M01.7, M02.4, M11, M1.7 and M2.4 models. We note that the majority of the BH mass estimates for the sources in our sample are measured using the continuum luminosity and the width of specific emission lines in single epoch spectra. Those measurements are thought to have systematic errors of the order of $\sim 0.3$ dex. Consequently, \ledd\ is also uncertain, even more so because \lbol\ is also uncertain as it is based on the use of bolometric correction factors, which have errors themselves. There is some additional uncertainty on the model time lags due to the error of the 2-10 keV luminosity as well (as parameters $m$ and $q$ in the \citet{lusso2012bolometric} equations also have errors). Assuming that, on average, \mbh, \ledd, and 2-10 keV luminosity are uncertain by a factor of $\sim 2$, then we estimate that the relative error on the model time lags can be up to 10-20 per cent (using the model time-lag plots in the Appendix of K21a). For that reason, we increased the error of $\bar{t}_{\rm lag,obs}$ by a factor of 1.1 when fitting the model to the observed time lags. 

Our results are listed in Table \ref{tab:results}. The first two columns list the various models we considered. The third column lists the best-fit corona height for each model. The respective $\chi^2_{\rm min}$ values are listed in the last column. Assuming that the corona height is the only free model parameter (for a given \spin\ and \fcol), then the number of degrees of freedom (dof) is 5, and only models M01.7 and M12.4 can fit the data well ($p_{\rm null}=0.04$ for these models). The null hypothesis probability, $p_{\rm null}$, is smaller than 10$^{-5}$ for all other models. 

\begin{figure}[t]
    \centering
    \includegraphics[width=\linewidth]{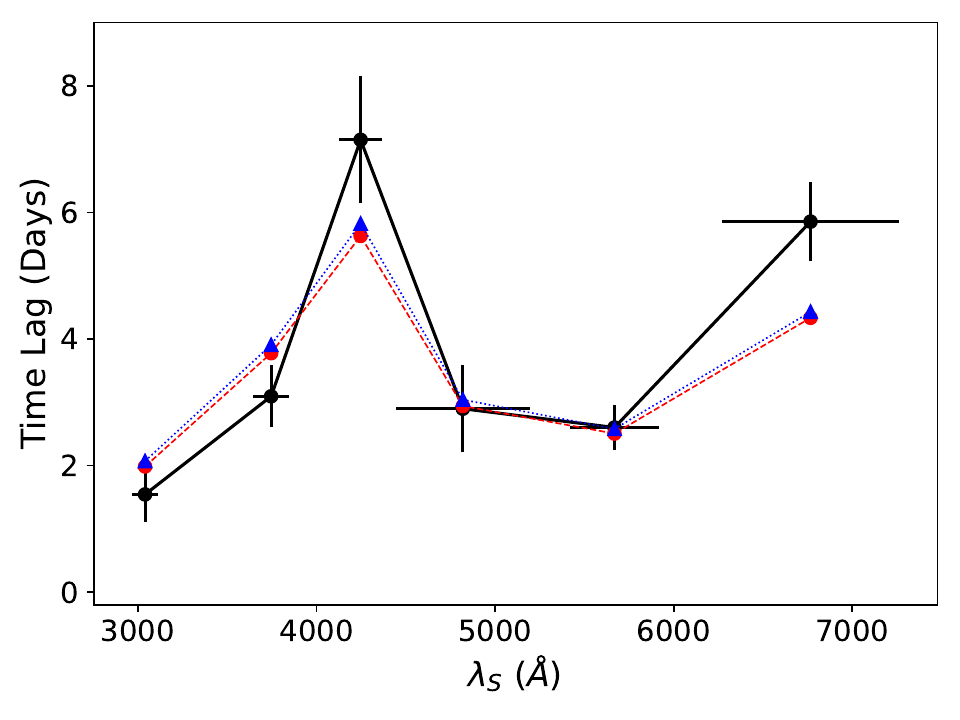}
    \caption{The observed, mean time-lags spectrum (black circles, connected with a solid line) and the best-fit, M01.7 and M12.4 models (red and blue points, respectively, connected with dashed lines).}
    \label{fig:bestfit}
\end{figure}

Figure \ref{fig:bestfit} shows the best fit results. Black filled circles connected with the solid line show \tlagobs\ plotted as a function of $\bar{\lambda}_{S}$ (their error is equal to the error of the mean multiplied by 1.1, as explained above). The red circles and the blue triangles, connected by the red and blue dashed lines, show the M01.7 and the M12.4 best-fit models, respectively. Both models fit well the data. 

\begin{figure}[t]
    \centering
    \includegraphics[width=\linewidth]{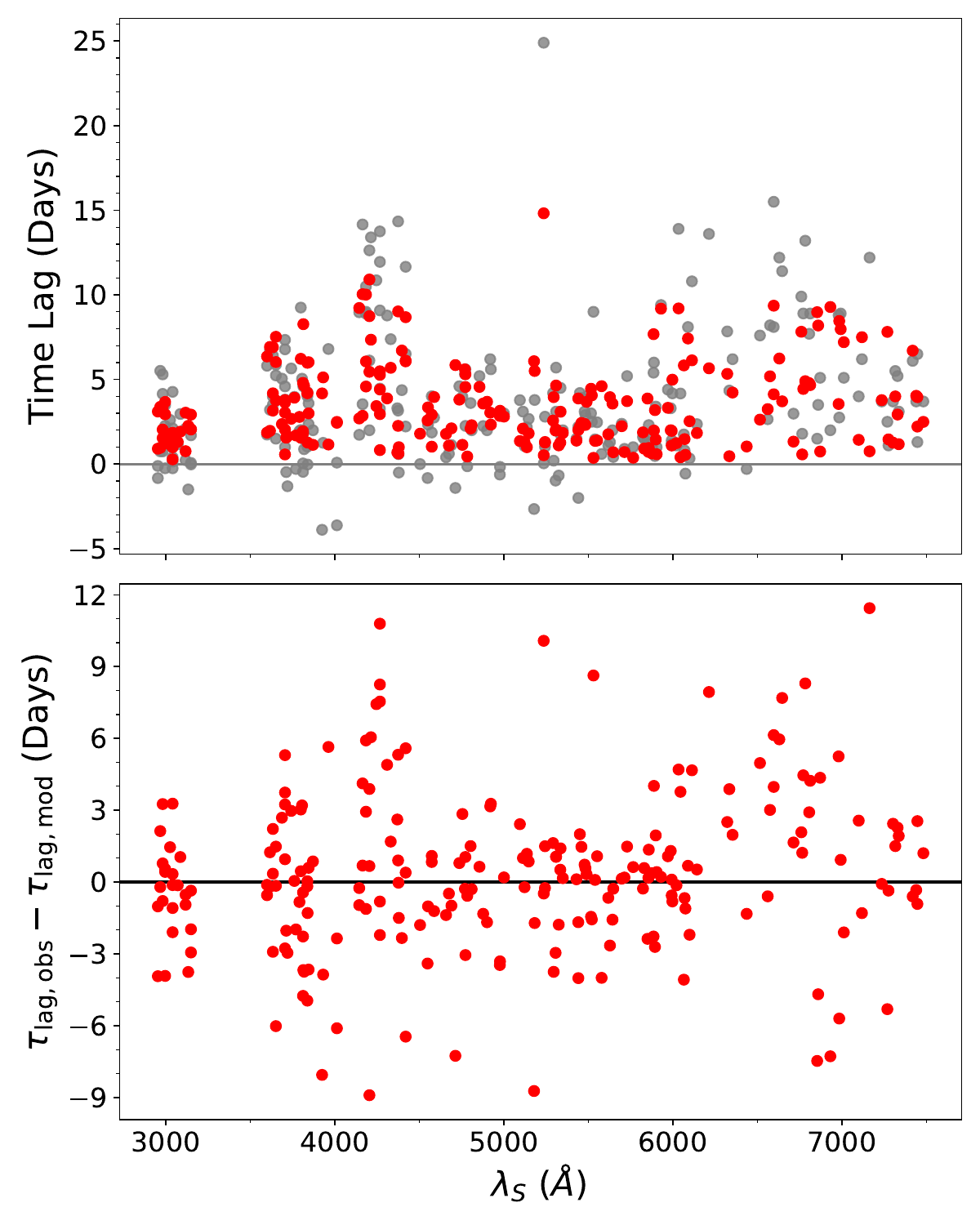}
    \caption{{\it Upper panel}: The observed and best-fit model time lags (for model M01.7) for the individual AGN in our sample (grey and red circles, respectively). {\it Bottom panel}: The respective best-fit residuals (i.e. observed minus the model predicted time lags).}
    \label{fig:modelobstlags}
\end{figure}

As a further indication that the models fit well the data, the upper panel in Fig. \ref{fig:modelobstlags} shows the observed time lags of the individual AGN  plotted as a function of $\lambda_{S}$ (grey filled circles), and the respective model time lags for M01.7, assuming the best-fit value for the corona height of 42 $R_{\rm g}$ (red filled circles). The bottom panel in the same figure shows the best-fit residuals (i.e. observed - model time lag). The data points in both panels in Fig.\,\ref{fig:modelobstlags} show that the model time lags match well, and can fully explain the range of the observed time lags. 

\begin{figure}[t]
    \centering
    \includegraphics[width=\linewidth]{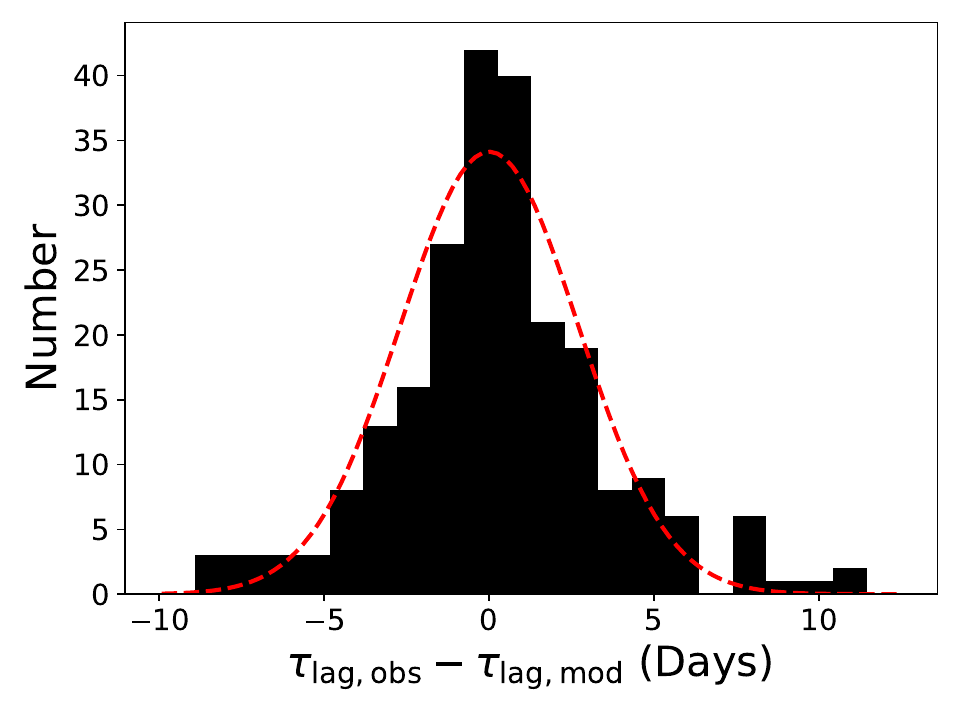}
    \caption{The distribution of the best-fit residuals plotted in the lower panel of Fig.\,\ref{fig:modelobstlags}. The red-dashed line indicates a Gaussian distribution with zero mean and standard deviation equal to the mean error of the time-lag measurements we used in this work.}
    \label{fig:residdistr}
\end{figure}

Figure \ref{fig:residdistr} shows the distribution of the best-fit residuals of the individual AGN, which are plotted in Fig.\,\ref{fig:modelobstlags}. The mean of the best-fit residuals is $0.36\pm 0.22$ days, which shows that it is consistent with zero, as expected if the model fits the data well. Furthermore, the standard deviation of the residuals is 3.3 days. This is very similar to the mean time-lag error, which is 2.7 days (this was computed using the errors reported in the papers we used to collect data, for the ICCF time-lag measurements). We fitted the distribution of the residuals shown in Fig.\,\ref{fig:residdistr} with a Gaussian of zero mean, and $\sigma=2.7$ days, using the Kolmogorov-Smirnov test. The red dashed line in the same figure indicates this Gaussian distribution. The K-S statistic was 0.074, which implies $p_{null}=0.15$, that is the distribution of the residuals is entirely consistent with a Gaussian distribution with mean zero, and standard deviation equal to the mean of the time-lag errors, as expected if the model fits the data well. This result further supports the hypothesis that the X-ray reverberation is fully consistent with the observed time lags.

Figure \ref{fig:modelobstlags}  shows that the best-fit residuals at wavelengths longer than $\sim 6200$\AA\, appear to be predominately positive, indicating that the model time lags may underestimate the observations at these wavelengths. This is also shown in Fig.\,\ref{fig:bestfit}, where the observed mean time lag at the longest wavelength is $\sim 2\sigma$ larger than the model prediction. Although this difference is not statistically significant we note that this is not surprising. According to K21a, their model time lags are valid for time-lag measurements at wavelengths shorter than 5000\AA. For certain combinations of the model parameters, the disc may be hot enough to contribute to the disc response even at radii larger than 10$^4$ $R_g$. Since this is the largest outer disc radius that K21a considered, the model time lags could be smaller than the observed ones at long wavelengths, if the disc radius is larger than 10$^4$ $R_g$. This could explain the slight discrepancy between the observed and model time lags that we observe at $\lambda \ge 6200$\AA.

\section{Summary and discussion}\label{sec:Sum_Disc}

We have used time-lag measurements for 128 AGN, which are more luminous, and host BHs which are more massive than in nearby AGN, and we measured their rest-frame UV/optical, average time-lags spectrum. We assumed the BH mass and bolometric luminosity estimates that are listed in the papers we used to collect data, we assumed an X-ray luminosity which is typical of Type I AGN (as determined by \cite{lusso2012bolometric}), and we fitted the mean time lags with the analytical, X-ray reverberation models of K21b, taking also into account the colour correction added later on by \cite{kammoun2023revisiting}. 

Our main result is that the observed time lags in quasars are fully consistent with the X-ray reverberation hypothesis. In particular, we find that the hypothesis of both spin zero and maximum spin for the BH in AGN can fit the data, as long as the colour correction factor is 1.7 and 2.4, respectively. To the best of our knowledge, this is the first time that the observed, continuum, UV/optical time lags in quasars are shown to be explained by X-ray reverberation of standard, Novikov-Thorne discs. 

Our results are also consistent with the \cite{papadakis2022x} results, who studied the half-light radius versus BH mass relation in AGN,  when the disc is illuminated by the X-ray corona. Contrary to past statements that the accretion disc size in quasars appears to be systematically larger than the standard thin disc model predictions by a factor of $\sim$2-4, \cite{papadakis2022x} showed that, when all the physical processes are taken into account, the half-light radii of X-ray illuminated discs are fully consistent with observations of  gravitationally lensed quasars. To the best of our knowledge, X-ray reverberation is the only model that has been shown to fit well both the observed luminosity and accretion disc size in gravitationally lensed quasars and the optical/UV, mean time lags in quasars. 



\subsection{Comparison with previous work}
\cite{jiang2017detection}, \cite{guo2022active} and \cite{jha2022accretion} found that the time-lag predictions provided from the X-ray illumination of the standard thin disc model were smaller by a factor of around $~2-3$ than the the observed ones. \cite{homayouni2019sloan} also found similar results, if they would consider only the high S/N objects in their sample or only the ones with positive time lags. 
The results from these studies are similar to results that have been reported from the study of the continuum UV/optical time lags from the recent, multi-wavelength, intense monitoring campaigns of nearby Seyferts, when the time lags were fitted by less accurate, X-ray reverberation model time-lag relations \citep[see e.g.][]{fausnaugh2016space, cackett2018accretion, edelson2019, hernandez2020intensive}

Actually, it was mainly because of the apparent disagreement between the X-ray reverberation models and the observed time lags that various other models were developed in order to explain the 
observations. For example, it was pointed out that reprocessed continuum emission originating in the broad-line region (BLR) gas could contribute to the wavelength-dependent continuum delays measured in AGN disc reverberation mapping experiments \citep[e.g.][]{Korista2001, Lawther2018, Korista2019, Netzer2022}. Such contribution should be expected and it is clearly evident in some \citep[e.g.][]{cackett2018accretion}, but not in all sources. In our case, the mean time-lags spectrum does not show any indication of additional contribution from the BLR gas. The diffuse continuum contribution to the time-lags spectrum should be significant at wavelengths below $\sim 3500\AA$, but we find no significant discrepancy between the model X-ray reverberation and the observed time lags at wavelengths smaller than $\sim 4000$\AA\ (see Fig.\, \ref{fig:bestfit}). 

Furthermore, alternative models were also developed in order to provide solutions to the (supposedly) ‘large-lags' accretion disc problem. For example, \cite{Hall2018} studied the case when the atmospheric density in the accretion disc is sufficiently low, in which case the disc flux in the UV/optical wavelengths is lower and the discs appear to have larger size when compared to discs that emit as blackbodies (for fixed bolometric luminosity). Their model can match the AGN STORM observations of NGC5548, but produces disc spectral energy distributions that peak at shorter wavelengths than observed in luminous AGN in general. \cite{Starkey2023} assumed a steep accretion disc height rim or rippled structures irradiated by the central X-ray source, and fitted successfully the NGC5548 time-lags spectrum, when assuming the steep disc height rim, or multiple crests, are located at a distance $\sim$ 5 light days away from the X-ray source. We cannot comment on whether such modifications to the standard picture of the central engine in AGN (i.e. a compact X-ray source illuminating a standard,  NT73 disc) are necessary because X-ray reverberation of canonical, NT73 accretion discs can fit the time lags we study. As we repeatedly shown in the past that X-ray reverberation can fit well the observed time lags in many nearby AGN as well (e.g. \citealp{kammoun2019hard}, K21b, \citealp{kammoun2023revisiting}). In fact, the X-ray reverberation model can fit the observed time lags for a large range of X-ray height and accretion rate values. 

Most of the previous studies that we have referenced in the Introduction, have used analytical X-ray reverberation time-lags models which do not depend on the source height, the spin of the BH, they do not take into account relativistic effects, as well as the X-ray flux that is absorbed depending on the disc ionisation state. All these effects are taken into account by the X-ray reverberation models that we use. Perhaps the most important difference, which affects significantly the model predictions for the time lags, is that previous models assumed that the ratio of the disc heating due to X-ray absorption over the disc internal heating due to accretion is constant with radius. We suspect that the inability of the previous models to account for the observed time lags is due to the fact that this assumption is not correct. The K21b analytic models that we use in this work were developed using the K21a disc response functions, which were calculated by computing the external to internal heating at each disc radius, for any combination of the other physical parameters (i.e. BH mass, accretion rate, height, X-ray luminosity etc). 


\subsection{The corona height and the BH spin in quasars}

Our results are based on the assumption that the X-ray source height is the same in all of AGN. We suspect this is not the case, and that there should be a distribution of heights in AGN. Our results then imply that, on average, the corona height is the same in all quasars, and it does not depend on parameters like the BH mass and/or the accretion rate. Interestingly, \cite{papadakis2022x} found that the height of the X-ray source should be larger than $\sim 40$$R_g$ in order to explain the observed half light radii in micro-lensed quasars. This is exactly what we also find in our work. The corona height should be equal or larger than 40 $R_g$ in order to explain the observed mean time lags in quasars. The fact that our model can explain these two observables, with the same model parameters, supports the hypothesis of X-ray thermal reverberation in AGN. 

We note that the corona height depends on the model assumption that the disc is flat. In reality, even if $H/R$ is roughly constant in accretion discs (where $H$ is the disc height at radius $R$), we expect $H$ to increase with increasing $R$. This will alter the incident angle of X-rays to the disc. It is possible then that X-ray corona with smaller heights will be able to fit the data equally well in this case. We plan to explore this possibility in the near future.

The X-ray reverberation time lags also depend on the BH spin. This was already discussed by K21a (see for example the difference in time lags for non-rotating and maximally rotating BH in their Figs. 16-23). It is for this reason that the analytical time-lags models of K21a and K21b are different for these two cases. Our results are based on the assumption that BH spin is the same in all AGN. Similarly to what we discussed above about the corona height, we expect there should be a distribution of spin parameters in AGN. Our results suggest that spin probably does not depend on other physical parameters of the system (like BH mass, or even redshift, perhaps, at least at redshifts smaller than 1). We find that both the hypothesis of zero spin and that of a maximally spinning BH can fit well the mean time lags.  

We have also assumed that \fcol\ is the same in all quasars. The best-fit model for \spin=1 is the one where \fcol=2.4. This value is appropriate for highly accreting sources \citep{ross1992spectra} and, since \ledd\ is lower than $\sim 0.4$ for most of the sources in our sample, our results suggest that spin should not be equal to one for most quasars. On the other hand, the best-fit model for spin zero is the one with \fcol=1.7. This is very similar to \fcol$\sim 1.71$, which is predicted by \cite{davis19} for the annulus with the highest effective temperature in the disc for \mbh$=2.6\times10^8$M$_{\odot}$ and \ledd=0.18 (see their equation 10, for a viscosity parameter of $\alpha=0.1$). Therefore, the spin zero solution is certainly accepted, but this does not necessary imply that BHs in quasars are indeed non rotating. Our results do not exclude a non-zero spin solution, they simply appear to exclude the possibility of a maximally rotating BH in most sources in the sample. 

The analytical models of K21b were developed just for two spin values, and the colour-correction factor applies to the full disc. We plan to use the codes of \cite{kammoun2023revisiting} in a future work, in order to investigate in more detail the constrains that the existing time-lag data could impose on BH spin in quasars, as well on the colour-correction factors. It is not certain that \fcol\ for the full disc will be well-approximated by the annulus with the highest effective temperature,  and the use of \cite{kammoun2023revisiting} could help us investigate this issue as it incorporates the colour-correction prescription of \cite{done2012fcol}, which does not imply the same \fcol\ for the full disc.

\begin{acknowledgements}
      We thank the referee for the helpful comments. This research has made use of the Spanish Virtual Observatory (\url{https://svo.cab.inta-csic.es}) project funded by MCIN/AEI/10.13039/501100011033/ through grant PID2020-112949GB-I00. This work utilises Matplotlib (\citealp{hunter2007matplotlib}), NumPy (\citealp{harris2020array}) and SciPy (\citealp{2020NatMe..17..261V}).
\end{acknowledgements}

%
%

\bibliographystyle{aa}
\bibliography{references.bib}

\end{document}